\documentclass[12pt]{article}
\pdfoutput=1
\usepackage{jcappub}
\usepackage{amsthm}
\usepackage{graphicx}
 \usepackage{amsmath}
\usepackage{amssymb}
\usepackage{color}
\usepackage{graphicx,textcomp,float,gensymb,wrapfig, enumitem,comment,dsfont,framed,slashed,appendix,wrapfig,xcolor}
\usepackage{ bbold }
\usepackage[margin=1in]{geometry}
\usepackage{braket} 
\usepackage{ mathrsfs }
\usepackage[small]{caption}
\usepackage{subcaption}
\usepackage{etoolbox,hyperref,makecell}
\patchcmd{\abstract}{\null\vfil}{}{}{}
\newcommand{\bea}{\begin{eqnarray}}  
\newcommand{\eea}{\end{eqnarray}}  
\title{Saving Natural Inflation}
\author{Djuna Croon and Ver\'onica Sanz}

\affiliation{
Department of Physics and Astronomy, University of Sussex, Brighton BN1 9QH, UK
}

\abstract{
Slow-roll inflation requires the inflaton field to have an exceptionally flat potential, which combined with measurements of the scale of inflation demands some degree of fine-tuning. Alternatively, the flatness of the potential could be due to the inflaton's origin as a pseudo-Goldstone boson, as in Natural Inflation. 
Alas, consistency with Planck data places the original proposal of Natural Inflation in a tight spot, as it requires a trans-Planckian excursion of the inflaton. Although one can still tune the renormalizable potential to sub-Planckian values, higher order corrections from quantum gravity or sources of breaking of the Goldstone symmetry would ruin the predictivity of the model. In this paper we show how in more realistic models of Natural Inflation one could achieve inflation without a trans-Planckian excursion of the field. We show how a variant of Extra-natural inflation with bulk fermions can achieve the desired goal and discuss its four-dimensional duals. We also present a new type of four dimensional models inspired in Little Higgs and Composite Higgs models which can lead to sub-Planckian values of the inflaton field.}

\emailAdd{d.croon@sussex.ac.uk}
\emailAdd{v.sanz@sussex.ac.uk}


\begin{document}
\maketitle
\flushbottom

\section{Introduction}

The idea of Inflation~\cite{inflation-idea} is a very successful paradigm, capable of explaining the cosmological data with the assumption that one field, the inflaton, dominates the evolution of the Universe in its early stage. Natural Inflation (NI) was first suggested in 1990 by Freese, Frieman and Olinto~\cite{here}, in answer to the hierarchy problem of inflation. This hierarchy problem addresses the fact that the condition for sufficient inflation combined with the amplitudes of the CMB anisotropy measurements imply that the width of the potential must be much larger than its height. Such a flat potential is generally considered unstable under radiative corrections, unless protected by some symmetry. 

In Natural Inflation a shift symmetry is invoked to protect the flatness of the inflaton potential. The inflaton that possesses this shift symmetry is an axion, a Nambu Goldstone boson from a spontaneously broken Peccei-Quinn symmetry. But as the inflaton potential cannot be fully flat, the shift symmetry cannot be exact. An explicit symmetry breaking is introduced which generates a potential for the axion, now a pseudo-Nambu Goldstone boson (pNGB). 
In its simplest form, natural inflation has the potential ~\cite{here}
\begin{equation} \label{VNI} 
V( \phi ) =  \Lambda ^4 \left(1 + \cos \left(\frac{\phi }{f}\right)\right)  .
\end{equation} 
Naturalness requires the spontaneous symmetry breaking scale (parametrized by the axion decay constant $f$) should be sub-Planckian, such that corrections from quantum gravity are suppressed. 

Since the advent of the cosmological precision measurements by the WMAP~\cite{wmap} and Planck~\cite{Planck} satellite, the bounds on the spectral index $n_s$ and the tensor to scalar ratio $r$ have significantly improved. The original natural inflation model now needs trans-Planckian scales $f$ to satisfy the most recent bounds, and therefore loses part of its motivation. A trans-Planckian decay constant $f$ renders the out of the range of validity of the effective theory. Indeed, albeit the Natural inflation potential requires a well-behaved potential
\bea
V_{NI} \ll \tilde{M}_P^4 \ ,
\eea 
this renormalizable potential will be corrected by non-renormalizable terms, e.g. 
\bea
V_{NR} = \phi^4 \, \left( \frac{\phi}{\tilde{M}_P}\right)^n \ , \label{NRpot}
\eea
where $\tilde{M}_p$ is the reduced Planck mass.
Under these circumstances, the success of this inflationary model depends on the ultra-violet (UV) completion of the theory. The origin of these non-renormalizable terms could be sources of breaking of the quantum gravitational effects, such as wormholes~\cite{wormholes} or any other source of breaking of the shift-symmetry at high-energies.

Models reconciling natural inflation with the CMB data should offer an explanation as to why the potential does not receive big gravitational corrections.  Different efforts have been made to do so but strive to achieve sub-Planckian values of the field, for instance, by considering a hybrid axion model~\cite{hybrid}~\cite{Peloso}, N-flation~\cite{N-flation}, and axion monodromy~\cite{axion-monodromy} and other pseudo-natural inflation models in Supersymmetry~\cite{pseudonat}.

One of the proposals to improve Natural Inflation relies on a potential generated by a Wilson line in extra-dimensional models, also known as Extra-natural inflation~\cite{extra-natural}. In these models, the inflaton is the fifth component of a gauge field in the extra dimension, and is thus protected from both higher-order corrections and gravitational effects by its locality in the extra dimension.
The potential is generated as 
\begin{equation} V ( \phi) = \Lambda^4 \sum_{I} \left[ (-1)^{F_I} \sum_{n=1}^{\infty} \frac{\cos \left( n \, q_I \frac{\phi}{f}\right)}{n^5} \right] \end{equation}
where $F_I = 1 \,(0)$ for massless bosonic (fermionic) fields with charge $q_I$. Here $f$ is the effective decay constant $f = 1/ (2 \pi g_4 R)$, with $R$ the size of the extra-dimension and $g_4$ the four-dimensional (4D) gauge coupling defined by $g_4^2 = g_5^2/(2 \pi R)$. 
In this scenario, loops of gauge bosons lead to the same form of potential as in the original model of natural inflation but with a different interpretation of the decay constant. Nevertheless, cosmological data indicates that $f > M_p $, leading to the slow-roll conditions  
\bea \label{feff} 2 \pi g_4 M_p R \ll 1 \eea
which requires a value of $g_{4D} \ll 1/2 \pi$ for a compactification scale below the Planck mass.

In this case Extra-natural inflation makes predictions for $n_s$ and $r$ very similar  from the predictions from Natural Inflation, as the higher terms in the sum are surpassed by $1/n^5$. These terms do become significant in the higher order slow-roll conditions, as was recently pointed out by~\cite{extranat2}, however, as both models predict values for the higher derivatives of the potential $V^{(III)}$ and $V^{(IV)}$ far below the current experimental limits, it is not possible to distinguish them yet.  

Here we will propose a mechanism to keep Extra-natural inflation within the validity of the effective theory by adding bulk fermions in the effective potential. Besides gauge bosons (with charge $q=1$)  we also consider the effects of fermions of fractional charge: up-type fermions of charge $+2/3$, and down-type fermions of charge $-1/3$. We will show that for certain combinations of gauge bosons and fermions the model can be made compatible with the Planck data for $f \leq \tilde{M}_p$. 

As a next example of a pNGB playing the role of the inflaton, we will consider a Coleman-Weinberg type potential generated by gauge and Yukawa couplings. We discuss its general structure, and study it numerically for the specific  form
\begin{equation} V ( \phi )  = \Lambda^4 \left(  \cos \frac{\phi}{f}  - \tilde\beta \sin^2 \frac{\phi}{f} \right) \end{equation}
This potential resembles the Minimal Composite Higgs model (MCHM) and the inflationary potential studied in~\cite{multinatural} upon choosing $ f_2 = f_1/2$ and $\Lambda_2 /2 = \Lambda_1$. 
We show that this model can be made compatible with the data for particular values of $\tilde\beta$. In the interpretation of $\phi$ as a pNGB, this corresponds to a relation between its couplings to fermions and bosons. 

Lastly we will consider the effects of quantum gravity and other UV breaking effects in the original model of Natural Inflation. We will consider  effective higher order operators, which we parametrize these as
$$ V( \phi ) =  \Lambda ^4 \left[1 + \cos \left(\frac{\phi }{f}\right) +  \sum_{n=5} c_n \, \phi^4 \, \left(\frac{\phi}{\tilde{M}_P}\right)^{n-4} \right] \label{HDOs}$$
We will investigate how these operators will affect the predictions of the model and show how the effect of tiny values of $c_n$ is able to bring the model outside the Planck region. This is an illustration of how the predictions of inflationary models when $f>\tilde M_p$ are lost unless one specifies very precisely the UV structure of the model.

The paper is organized as follows. We present the general set-up for the inflaton as a pseudo-Goldstone boson in Sec.~\ref{generalsetup}, moving to discuss the origin of the original Natural Inflation model, which we call {\it Vanilla Natural Inflation}, and its clash with Planck data in Sec.~\ref{vanilla}.  We introduce a variant of the Natural Inflation scenario, namely Extra-Natural Inflation and explain how it leads to a similar clash unless one introduces bulk fermions, see Sec.~\ref{sec:extranat}. In Sec.~\ref{sec:4DCW} e finally move onto purely four-dimensional models inspired in the ideas of Little Higgs and Composite Higgs, where the pseudo-Goldstone is generated via Coleman-Weinberg with gauge and Yukawa contributions. 

\section{The general set-up of Natural Inflation}\label{generalsetup}

The idea that the inflaton is a pseudo-Goldstone boson could explains the flatness of the potential required to generate enough inflation, without resorting to fine-tuning.
The basic ingredient of all models of natural inflation is the existence of a approximate global symmetry, broken spontaneously at high energies. Hence, one expects a new degree of freedom, a pseudo-Goldstone boson (pGB). \bea
\Phi \to \Phi + C \label{shiftsym}
\eea
where $C$ is a constant.

In this scenario, the pGB inflaton is originated by the breaking of a {\it global} symmetry $G$ to a subgroup $H$. We will denote the generators of $G$ as $T^a \in H$ and $T^{\hat{a}}$), where $T^{\hat{a}}$ are the broken generators. The resulting pNGBs are associated with these broken generators, and we will assume the inflaton field is one linear combination of them.  The mass and couplings of the pGB would depend on how the global symmetry is explicitly broken.

 Note that one can express the Goldstone boson as
\bea
\Sigma (x) = \Sigma_0 e^{i \Pi(x)/f} \textrm{ where } \Pi = T^{\hat a} \, \Phi^{\hat a} \label{phisigma}
\eea

As $\Phi^{\hat a}$ appears in exponential form, its potential will be forbidden by the shift symmetry,  Eq.~\ref{shiftsym}. The inflaton is then a linear combination of  the pseudo-Goldstone bosons, and for large symmetry groups there could be more than one inflaton. In this paper, we will discuss a simple scenario with only one inflaton but it could be generalized to variants of hybrid inflation.

The potential for the inflaton could be generated in several ways, leading to different predictions for inflation. We will consider various options in this paper, namely
\begin{itemize}
  \item A gauge group, external to $G$ and $H$, breaks the symmetry $G$. The archetypical example is instanton effects and explicit breaking through quark mass terms, as in models for axions (QCD or hidden).   
    \item An extra-dimensional gauge theory breaks downs spontaneously via compactification, leading to extra-dimensional components of the gauge field exhibiting a shift symmetry. The potential is then generated as a non-zero expectation value of a Wilson line of the inflaton field, due the explicit breaking via Yukawa and new gauge couplings. This is the proposal of  Extra-natural inflation. We will discuss how the extra-dimensional model has a dual description in terms of purely four-dimensional models. 
  \item In theories where the inflaton is a 4D Goldstone boson, instead of relying on non-perturbative instanton effects to generate the potential, one could weakly gauge some of the global symmetries and also consider explicit breaking through Yukawa couplings. The potential is then generated as a Coleman-Weinberg contribution from fermions and gauge bosons. This is a popular mechanism to build Little Higgs and Composite Higgs models~\cite{LH, Composite-Higgs-orig}.
\end{itemize}

In this paper we will not consider the situation of the inflaton as a pseudo-Goldstone boson of the spontaneous breaking of a space-time symmetry, such as the dilaton or the radion in extra-dimensions, see Ref.~\cite{John} for some work along these lines.

\section{Vanilla Natural Inflation: Instanton-like potential}~\label{vanilla}

In this section we introduce the basic idea of Natural Inflation, with a potential generated through instanton effects.  We will then discuss the current situation of these models when confronted with cosmological data.

Consider a general non-abelian gauge theory $X$. Instantons are solutions to the gauge equations of motion
\bea
D_{\mu} F^{\mu\nu} = 0 \textrm{ at } |x|\to \infty \textrm{  with } A_{\mu} \to U \partial_\mu U^\dagger  \ .
\eea  
Instanton solutions can be found when there is a non-trivial element $U(x)\in \pi_3 (G)$, the third homotopy group of $X$. These solutions are characterized by the size of the inflaton $\rho$ and the number of possible orientations under the gauge group, e.g. for $SU(N)$ there are $4 N$ orientations. The classical action of an instanton solution is given by
\bea
S_{inst}= \frac{8 \pi^2}{g^2} = \frac{2 \pi}{\alpha} \ .
\eea
Instanton effects are non-perturbative, the suppression $e^{-S_{inst}}$ selects gauge sectors with $\alpha \sim {\cal O}(1)$. Note that instantons here are treated more generally than QCD instantons; they could be world-sheet, membrane instantons or supersymmetric instanton effects.

Assume that the inflaton is a singlet of the symmetry $X$ and couples to the $n$-instanton solution through a term respecting the shift symmetry, Eq.~\ref{shiftsym}. Then, 
\bea
V_{n-\Phi}=-\Lambda^4 \,  e^{-S_n} \, \Sigma^n
\eea

But as we must also consider the effect of $n$-anti-instanton solutions and sum over $n$,
\bea
\sum_{n=1,\infty} V_{n-\Phi} + V_{\bar n-\Phi} \simeq 
 -\Lambda^4 \,  e^{- \frac{2 \pi}{\alpha}}  (\Sigma+\bar \Sigma) = -2 \Lambda^4 \,  e^{- \frac{2 \pi}{\alpha}} \cos(\Phi/f) \label{Snphi}
\eea
the potential generated by instantons becomes of the familiar $\cos (\Phi/f)$ form.
This form of the solution is independent of the origin of $X$, as all instanton solutions adopt the same form as a $SU(2)$ instanton~\cite{coleman-aspects}. 

Here we have taken $\Lambda$ as the scale which allows the inflaton to couple to the instantons. For example, for QCD instantons, $\Lambda$ is related to the QCD pion sector via the fermion trace anomaly, $\Lambda^4 \simeq f_\pi^2 m_\pi^2$.  

Additional symmetries in the theory can make the instanton contributions to the inflaton in Eq.~\ref{Snphi} vanish. For example, in Supersymmetry (SUSY) the coupling of instantons to the inflaton appears through a superpotential $W$~\cite{witten}
\bea
W_{SUSY} = M^3    e^{-S_n} \, \Sigma \ .
\eea
No contribution with $\bar \Sigma$ appears in the superpotential, as it is holomorphic in the chiral superfield $\Sigma$. Thus under these circumstances there is no dependence on $\Phi$ in the potential generated by F-terms ($|\partial W/\partial \Sigma|^2$) neither in supergravity contributions ($\propto |W|^2$).  But this situation changes once SUSY is broken. For example, assume there is a non-zero $F$-SUSY breaking term, $F\sim M_{\slashed{SUSY}}^2$, then the interference of this source of SUSY breaking term and the inflaton's would lead again to a potential
\bea
V \simeq  M_{\slashed{SUSY}}^2 \Lambda^2  e^{- \frac{2 \pi}{\alpha}} \cos(\Phi/f) \label{Snphi} \ .
\eea

Finally we consider the situation that more than one instanton solution contributes to the inflaton potential. For example, if the gauge symmetry $G$ is broken down to another non-abelian gauge symmetry $H$, instantons from both theories could contribute to the inflaton potential. This situation again leads to a $\cos \Phi/f$ potential~\cite{csaki-terning}. 

We can use the invariance of the potential under a shift of $\Phi/f$ modulus $2 \pi$ to rewrite the solution as a potential with a minimum at $\Phi=0$, and we will do so in the next section.

\subsection{The problem with vanilla natural inflation}

The vanilla natural inflation model (Eq.~\ref{VNI}) is an example of slow-roll inflation; that is, it satisfies the conditions $\epsilon \ll 1$ and $\eta \ll 1$, where $\epsilon$ and $\eta$ are here given by 
\bea \epsilon = \frac{\tilde{M}_p^2}{2} \left( \frac{  V'(\phi)}{V(\phi)}\right)^2  \textrm{  and   } \eta = \tilde{M}_p^2 \frac{V''(\phi)}{V(\phi) } \ .
\eea

To simplify our expressions, in this section we work in units of reduced Planck mass  $ \tilde{M}_p$; that is, we will rescale our parameters $\phi \rightarrow \frac{\phi}{\tilde{M}_p}$   and  $f \rightarrow \frac{f}{\tilde{M}_p}$.

The number of e-foldings in the slow-roll approximation is then given by 
\bea N = \frac{1}{ \sqrt{2}} \int_{\phi_E}^{\phi_I} \frac{1}{\sqrt{\epsilon}} \eea
where $\phi_E$ is fixed as the field value for which either $\epsilon = 1$ or $\eta = 1$, in other words, the field value for which the slow-roll approximation breaks down. 

For a model with potential (Eq.~\ref{VNI}) we have 
\bea N = 2 f^2 \left[\log \left(\sin  \frac{\phi_E}{2 f} \right)-\log \left(\sin  \frac{\phi }{2 f} \right)\right] \eea
where 
\bea \phi_E = f \tan ^{-1}\left(\frac{1-2 f^2}{2 \sqrt{2} f}\right) \eea

Now solving for the field in terms of the number of e-foldings, we obtain 
\begin{equation} \label{phiNI} 
\phi_{NI} = 2 f \sin ^{-1}\left[\exp\left(\frac{2 f^2 \log \left(\sin \left(\frac{\phi_E}{2 f}\right)\right)-N}{2 f^2}\right)\right]
\end{equation} 

\begin{figure}[t] 
\centering
  \includegraphics[width= 300pt]{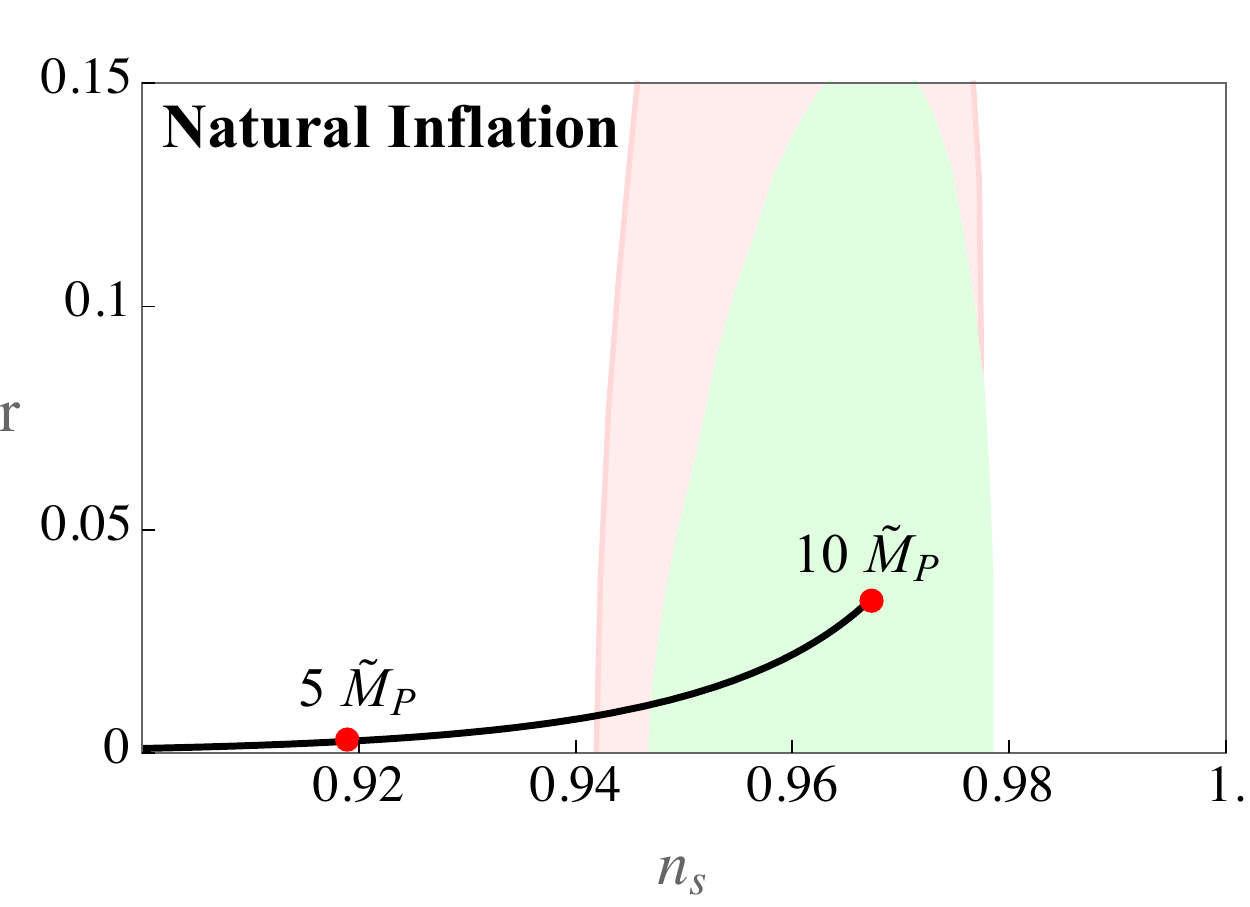}
  \caption{{\it Natural Inflation:} Values of cosmological parameters in the ($n_s$, $r$) plane with $N=60$. The red points correspond to   $f = M_p \simeq 5 \tilde{M}_p$, and $f = 2 M_p \simeq10 \tilde{M}_p$. The regions correspond to a combination of Planck, WP and BAO data, where the green (pink) region is the 95\% CL assuming the $\Lambda$CDM hypothesis and $r$ (and running of the spectral index). }\label{NIplot}
\end{figure}

In Fig.~\ref{NIplot} we show the values of $n_s$ and $r$ for this scenario, setting the number of e-foldings to $N=60$. We do not show the line for $N=50$, but it lies above the $N=60$ with similar values of $f$, see for example Ref.~\cite{Freese-Kinney}. The pink and green regions correspond to a fit of Planck, WMAP and baryon acoustic oscillations (BAO) to $n_s$ and $r$ with and without running of the spectral index~\cite{Planck}.

Successful vanilla natural inflation requires values of the decay constant $f\gg \tilde{M}_p$, hence any higher-dimensional operator of the type in Eq.~\ref{NRpot} ,
could dominate over the original renormalizable potential~\cite{xavier, wormholes}. In this situation, the inflationary theory ceases to be a good effective description, losing predictivity unless a complete understanding of the theory in the UV is achieved, including quantum gravity effects. 

Nevertheless, as this term violates the discrete shift symmetry of the original Lagrangian, one could devise a UV completion which (approximately) respects the symmetry, parametrically suppressing the dangerous terms. Besides, one could envision taming quantum gravity effects by  embedding  the shift symmetry in a gauge symmetry at high-energies~\cite{wormholes}. 

Also note that extending the axion-like inflationary scenario does not seem to circumvent the problem of trans-Planckian values of $f$. This is the case of axion monodromy~\cite{axion-monodromy}, where one obtains $f \gg \tilde{M}_p$, or even N-flation~\cite{N-flation}, where one trades a large $f$ by a large number ($\sim 10^3$)  of axion-like inflatons~\cite{burgess}.

Let us finish by discussing another way to study this inflationary scenario which does not rely on the $n_s$, $r$ plane. One can couple the inflaton to gauge fields~\cite{1410.3808v1,1110.3327v1},  
\bea\label{gaugeaxion} \Delta \mathcal{L} = - \frac{1}{4} C(\phi) \epsilon^{\mu \nu \rho \sigma} F_{\mu \nu } F_{\rho \sigma} \eea
as this transforms as a total derivative and therefore does not induce perturbative corrections to the inflaton potential: $ E \cdot B \sim \epsilon^{\mu \nu \rho \sigma} F_{\mu \nu } F_{\rho \sigma}  \sim \partial_\mu \left(  \epsilon^{\mu \nu \rho \sigma} A_{ \nu } F_{\rho \sigma} \right)$; this is reminiscent of the original use of the axions to solve the strong CP problem. 
The term Eq.~\ref{gaugeaxion} may give rise non-Gaussianity and gravitational wave signals, and provides a decay channel for the inflaton. 

Mild symmetry breaking effects may further give rise to the term~\cite{1110.3327v1}
\bea \Delta \mathcal{L} =- \frac{1}{4} B(\phi) F^2 \eea
However, slow-roll requires the effect to be very small, and thus $B(\phi)$ to be nearly constant. 

It was shown in Refs.~\cite{0908.4089v2,1110.3327v1} by using $H_{int} \sim E \cdot B$ in the mean field equations for $\phi$ that sufficiently large coupling to gauge fields causes a back-reaction. This is a purely classical effect, in which inhomogeneities in the inflaton field are sourced by those in the electromagnetic field. It increases the amount of inflation by about $10$ e-foldings, or, equivalently, changes the spectral index $n_s$ and the tensor to scalar ratio $r$ with the same amount of efoldings. For sufficiently strong coupling to a large number of gauge fields one could accommodate $f < \tilde{M}_p$ within the experimental bounds. For such a coupling the model would also predict observable (but currently within the bound) non-Gaussianity from inverse decay.

\section{Extra-natural inflation: a 5D model and its duals}~\label{sec:extranat}

In this section we will explore a different route to generate a pNGB inflaton potential, which does not rely on instanton effects.The inflaton's shift symmetry could be the remnant of an extra-dimensional gauge symmetry broken down by compactification~\cite{extra-natural}. Moreover, we will discuss how this extra-dimensional mechanism has dual descriptions in terms of purely four-dimensional (4D) models.

To show how this mechanism works, let us discuss a simple example in a five-dimensional (5D) flat spacetime (the analysis can easily be generalized to warped space-times). Consider a gauge field in 5D,
\bea
{\cal S} = -\frac{1}{4} \int d^5 x F_{MN} F^{MN}
\eea
where $M$, $N$ run over four-dimensional indexes $\mu=0-3$ and over 5, the index along the fifth dimension. Before compactification, the 5D gauge invariance is  given by
\bea
A_M (x) \to A_M (x) - \partial_M \alpha(x) \ .
\eea
The reduction from 5D to 4D can be done by compactifying the extra-dimension $x_5$ on an orbifold $S^1/{\mathbb{Z}_2}$ with $x_5 \in [0,L]$, by specifying  the boundary conditions of the gauge field on the orbifold fixpoints at 0 and $L$. Let us consider two choices consistent with the orbifold,
\bea
\textrm{\bf 4D gauge: } & & \partial_5 A_\mu =0 \textrm{ and } A_5 = 0 \textrm{ at } x_5 = 0, \, L \nonumber \\
\textrm{\bf Shift-symmetry: } & & A_\mu =0 \textrm{ and } \partial_5 A_5 = 0 \textrm{ at } x_5 = 0, \, L ,
\eea
where $\partial_5$ denotes the derivative along the extra-dimension. In the first case, the low energy theory exhibits a 4D gauge symmetry, whereas in the second case all the gauge bosons become heavy except a massless $A_5$ 4D zero-mode. The resulting low-energy theory of this zero-mode exhibits a shift symmetry, remnant of the 5D gauge symmetry. The $A_5$ can couple to any species charged under the 5D gauge group through a non-local gauge invariant Wilson line, $e^{i \oint_{0}^L A_5 }=e^{i A_5 L}$. 

Bulk fields propagating between the two orbifold fixed-points will radiatively generate a non-zero value for this Wilson line and hence provide a potential for $A_5$, our inflaton candidate. Indeed, charged fermions and gauge bosons would have non-trivial boundary conditions in the presence of $A_5$ in the spectrum. For example, the equation of motion of a fermion coupled to $A_5$ would be solved with modified boundary conditions $\Psi(x_\mu, x_5)=\Psi(x_\mu, x_5+L) \, e^{i e t\oint d x_5 A_5}$~\cite{japanese_wilson}. We can then obtain the contribution to the inflaton potential from fermion and gauge degrees of freedom from the bulk as a closed loop of fields propagating in the bulk, with opposite sign fermionic and bosonic contributions. These contributions are periodic on the inflaton field and also proportional to the charge of the field under the extra-dimensional gauge symmetry. As announced in the introduction, the potential in Extra-natural inflation then takes the form,
$$ V ( \phi) = -\Lambda^4 \sum_i (-1)^{F_i} \sum_{n=1}^{\infty} \frac{1}{n^5} \, \cos \left( n \, q_i \frac{\phi}{f}\right)  \ ,$$
where $F_i$ is the fermionic number of species $i$, and $q_i$ its charge. The inflationary potential scale is related to the size of the extra-dimension  $L=2 \pi R$ by the expression~\cite{extra-natural} 
\bea
 \Lambda^4  = \frac{3}{4 \pi^2} \frac{1}{L^4} \ ,
\eea
and the inflaton's decay constant is given by $f=1/(g_{4D} L)$, with $g_{4D}$ the 4D gauge coupling of the gauge group which generated $A_5$.

In the next section we will discuss viable models of inflation in this context. But before we move onto comparing with cosmological data, let us discuss the fact that these models can be viewed as four-dimensional models,  dual versions of the extra-dimensional model. In the 5D picture, the spontaneous breaking of the gauge symmetry by compactification acts as a Higgsing mechanism. Al low energies compared with the compactification scale (the mass of the Higgsed gauge bosons), the original gauge symmetry is realized as a global symmetry with $A_5$ a remaining Goldstone boson. 

{\bf Deconstructed dual: } One 4D dual of Extra-natural inflation is deconstruction~\cite{deconstruction}. Instead of a 5D model, one could consider a set of $N$ copies  of the same 4D gauge group linked through bi-fundamental fields, called {\it links}. The role of the inflaton in this case is played by the Goldstone mode contained in the link field $\Sigma_k$ system. Each link field is given by
\bea
\Sigma_k= e^{i \frac{\Pi_k}{\tilde f}}
\eea
and the inflaton is in this case
\bea
\Phi = \frac{1}{\sqrt{N}} \sum_{k=1}^N \Pi_k
\eea
Note that as the inflaton is a linear combination of $N$ fields, its decay constant is $f=\sqrt{N}\tilde{f}$ with $\tilde f$ the decay constant of each of the links $\Pi_k$. The potential generated in the deconstructed models is equivalent to the Extra-natural inflation but instead of a Wilson loop in 5D one computes a Coleman-Weinberg potential in this gauge configuration. For example, gauge contributions would lead to~\cite{deconstr-EWSB}
\bea
V(\Phi) = \frac{-9 }{4 \pi^2} \frac{g^2 f^4}{N^2} \sum_{n=1}^\infty \frac{\cos(2 \pi N n \Phi/f)}{n (n^2 N^2-1) (n^2 N^2-4)} \ .
\eea 

{\bf 4D Global symmetry dual: } Another way to view the Extra-natural inflationary model is the use holographic methods, based on the AdS/CFT correspondence~\cite{adscft}. In this approach, 4D global symmetries correspond to 5D gauge equivalents. Therefore, the 4D dual of $A_5$ is a 4D Goldstone boson resulting from the spontaneous breaking of a global symmetry $G$ down to $H$. This breaking is due to some new strongly coupled sector, and the 4D bound states from the strongly coupled sector are the duals of the Kaluza-Klein excitations in the Extra-natural inflationary model.

\subsection{The issue with the original Extra-natural inflation model, and how to solve it}

The original model of Extra-natural inflation~\cite{extra-natural} considered an inflaton potential generated by bulk gauge bosons. In this case the model makes predictions for $n_s$ and $r$ very similar to Natural Inflation, and therefore suffers from the same  issues. In the 5D picture $f > \tilde{M}_p$ may not be problematic (as long as (Eq.~\ref{feff}) is satisfied; that is, for sufficiently small 4D gauge coupling $g_4$), but in deconstructing dimensions it implies the problematic relation $f_{link} \gg \tilde{M}_p$ ~\cite{extra-natural}. 

\begin{figure}[t] 
\centering
  \includegraphics[width= 300pt]{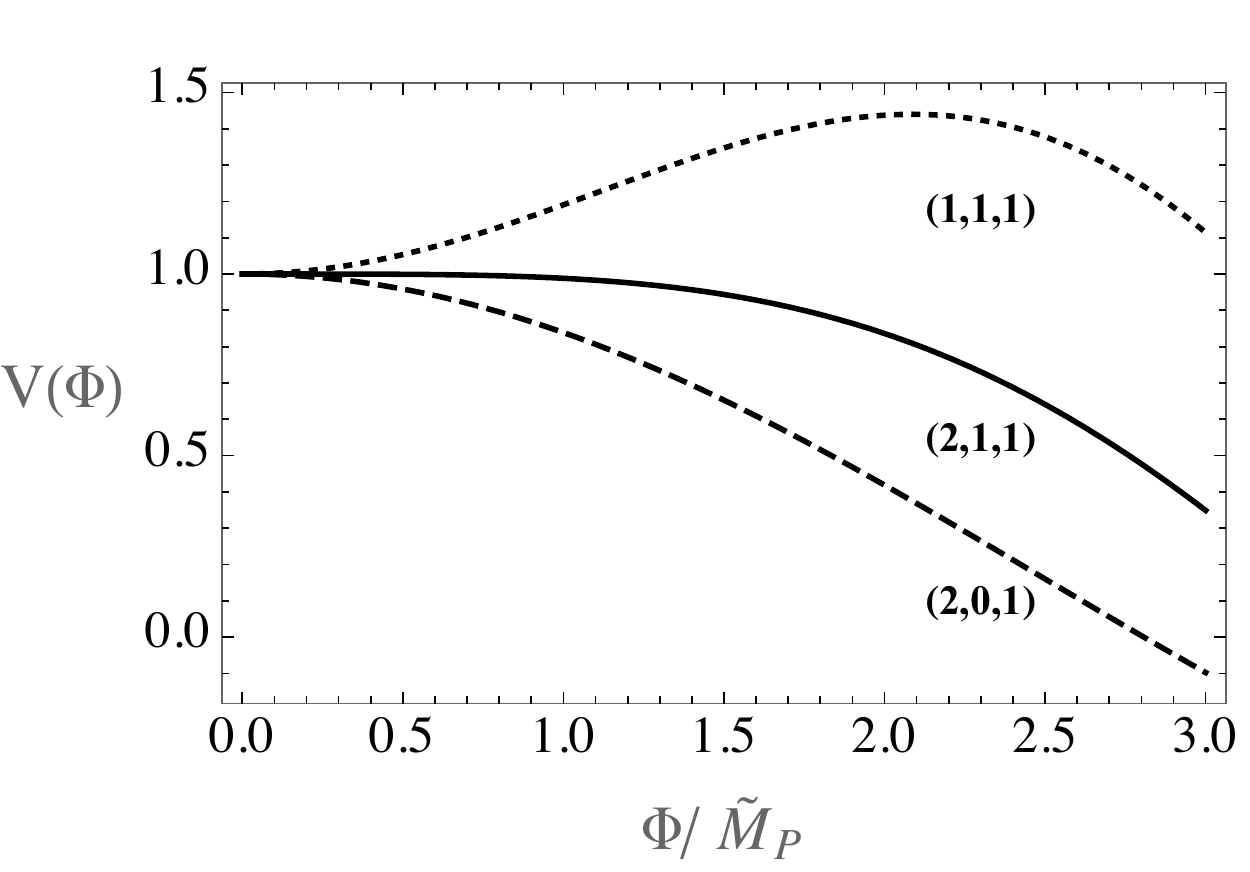}
  \caption{{\it Extra-natural inflationary potential: } Shape of the inflationary potential for different choices of $(\text{N}_U,\text{N}_D, \text{N}_V)$. Models with more bosons than fermions do not generate inflation around $\phi$=0.}
  \label{fig:wrongpot}
\end{figure}

Here we will investigate the effect of adding fermions to the original model. We consider gauge bosons (with charge $q=1$) and fermions of fractional charge (up-type fermions of charge $+2/3$, and down-type fermions of charge $-1/3$). The potential becomes
\bea\label{ENIpot} V (\phi) =  \Lambda^4 \left[ \text{N}_U \cos \left(\frac{2 \phi}{3 f}\right)+ \text{N}_D \cos \left(\frac{\phi}{3 f}\right) -\text{N}_V \cos \left(\frac{\phi}{f}\right) \right] \eea

We shall classify the different models using $(\text{N}_U,\text{N}_D, \text{N}_V)$. Considering non-integer charge fermions eludes the $\cos(\phi/f)$ form of the potential, which would lead to the same situation as in the vanilla inflation case discussed in the previous section.

We are specifically interested in the flat region around the origin of the potential: our setup (Eq.~\ref{ENIpot}) may be flatter than the vanilla model in this region. Close to $\phi=0$ the derivative of the potential is approximately given by
$$ V'(\phi) \approx \left( \text{N}_V - \frac{\text{N}_D}{9} -\frac{4 \text{N}_U}{9} \right) \phi \ , $$ hence when $\text{N}_U$ and $\text{N}_D$ are small compared to $\text{N}_V$, $V'(\phi)$ has the wrong sign at the origin, see Fig.~\ref{fig:wrongpot}. It is clear that one needs to consider models with fermions to render the inflationary theory viable.

\begin{figure}[t] 
\centering
  \includegraphics[width= 300pt]{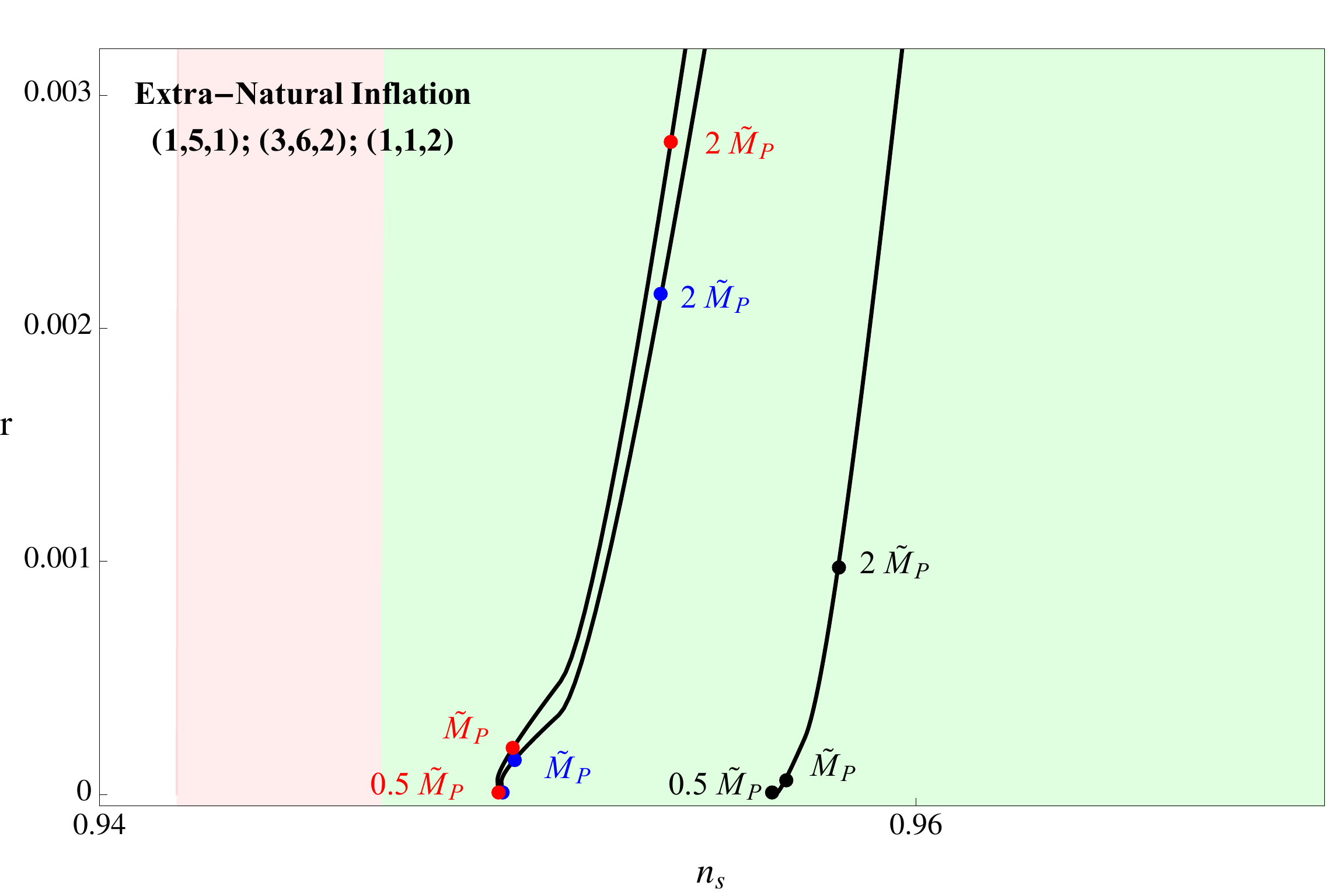}
  \caption{{\it Extra-natural Inflation:} Values of cosmological parameters in the ($n_s$, $r$) plane with $N=60$. The bullet points correspond to   $f = 0.4 M_p \simeq 2 \tilde{M}_p$, $f = M_p/5 \simeq \tilde{M}_p$ and $f = M_p/10 \simeq \tilde{M}_p/2$. The regions correspond to a combination of Planck, WP and BAO data, where the green (pink) region is the 95\% CL assuming the $\Lambda$CDM hypothesis and $r$ (and running of the spectral index).} \label{fig:extranaturalnsr}
\end{figure} 

Therefore, one can classify configurations which could lead to successful inflation using the condition near the origin, e.g.
\bea
(\text{N}_U,\text{N}_D, \text{N}_V) &=&  (0, 9, 1) \ , (1, 5, 1) \ , (2, 1, 1) \ , (2, 10, 2) \ , (3, 6, 2) \ , (4, 2, 2) \  , \nonumber \\ 
& & (5, 7, 3) \ , (6, 3, 3) \ , (7, 8, 4) \ldots 
\eea
In the cases we consider, the flatness of the potential guarantees that $r<{\cal O}(0.1)$. Therefore we concentrate on compatibility of the spectral index when confronting the models with the cosmological data. 
  
Not all these models would give successful inflation for a sub-Planckian decay constant $f$. Nonetheless, we do find successful inflation for a number of models, such as   $(\text{N}_U,\text{N}_D, \text{N}_V)= (2,1,1)$, (1,5,1), (3,6,2), (0,9,2) and others, including any integer number of times the values of $(\text{N}_U,\text{N}_D, \text{N}_V)$ mentioned above. 

In Fig.~\ref{fig:extranaturalnsr} we show how for these solutions, successful inflation with sub-Planckian values for the field can be obtained with very little variation in the value of $n_s$. We have checked that the higher-derivatives of the potential satisfy the  Planck bounds on $V^{(''')}$ and  $V^{('''')}$.

\section{Realistic 4D Inflation: Coleman-Weinberg Inflation}\label{sec:4DCW}

In this section we consider a new type of inflationary model, with the inflaton a four-dimensional pseudo-Goldstone boson, and a potential generated via Coleman-Weinberg contributions from the explicit breaking of the global symmetry by weak gauging and Yukawa couplings, as mentioned in section \ref{generalsetup}. Indeed, loops of gauge bosons and fermions will generate a Coleman-Weinberg potential for the inflaton, 
\bea
V^g_{CW} + V^\Psi_{CW} &=& 
\frac{N_V}{2} Tr  \int \frac{d^4 p}{(2 \pi)^4} \log (p^2+M_V^2(\Phi) ) \nonumber \\
&-& N_\psi Tr \int \frac{d^4 p}{(2 \pi)^4} \log (p^2+M_\Psi^\dagger(\Phi) M_\Psi(\Phi)) \label{CWf}
 \label{CWf}
\eea
where $M_V(\Phi)$ ($M_\Psi(\Phi)$) is the mass term of the gauge bosons (fermions) in the presence of $\Phi$ as a background field. $N_V$ is the number of vector degrees of freedom, $N_V=2$, 3 for massless (massive) vector bosons and $N_\psi$ is the number of Weyl fermions contributing to the potential. We will consider that our inflaton is a linear combination of the pseudo-Goldstone degrees of freedom in $\Phi$ and label it $\phi$.

In theories where $\Phi$ is generated as a Goldstone boson, one can write the couplings to fermions and gauge fields in a general way as
\bea
{\cal L} &=& \frac{1}{2} P^{\mu\nu} \, \left(\tilde \Pi_0 (p^2) Tr(V_\mu V_\nu)+\tilde \Pi_1 (p^2) \Sigma V_\mu V_\nu \Sigma^t \right) \nonumber \\
&+&  \bar \Psi \slashed{p} (\Pi_0 (p^2) + \Pi_1 (p^2) \Gamma_i \Sigma_i) \Psi  \label{PsiSigma}
\eea
where $\Sigma$ is given in Eq.~\ref{phisigma}, the fermion $\Psi$ is written in a representation of the group $G$, and $\Gamma_i$ are the fermionic Gamma matrices in this representation, and related to the broken generators $T^{\hat a}$. For example, see Sec.~\ref{MCHM} for the explicit form of these matrices in the breaking $G/H=SO(5)/SO(4)$. $P^{\mu\nu}$ is the inverse propagator, namely for a gauge boson $P^{\mu\nu}=\eta^{\mu\nu}- p^\mu p^\nu/p^2$.

The form of the mass terms $M_\Psi(\Phi)$ and  $M_V(\Phi)$ in Eq.~\ref{CWf} can be read from Eq.~\ref{PsiSigma} after specifying the type of breaking responsible for the emergence of the Goldstone bosons. Generally speaking, the mass term is also a periodic function of the Goldstones, that is, a combination of 
\bea
 s_{\Phi} \equiv \sin{\frac{\Phi}{f}}  \textrm{ and } c_\Phi \equiv \cos{\frac{\Phi}{f}}  \ ,
\eea
and therefore can be written in generality as 
\bea
{\cal L}&=& \bar \psi_i  \, \slashed{p} \, \left(\Pi_0 (p^2)_{ij} +  \Pi_1^{s} (p^2)_{ij} \, s_\Phi+ \Pi_1^{c} (p^2)_{ij} \, c_\Phi \right)\,  \psi_j \nonumber \\ 
&+& \frac{1}{2} P^{\mu\nu} \, \left(\tilde \Pi_0 (p^2) Tr(V_\mu V_\nu)+ (\tilde \Pi_1^{s^2} s^2_\Phi +\tilde\Pi_1^{sc} s_\Phi c_\Phi +\tilde \Pi_1^{c^2} c^2_\Phi )  V_\mu V_\nu  \right)
\eea
where the index $i$, $j$ run over all Dirac fermions.

In the following we discuss in detail the fermion case, as the vector case is a simpler variation of the same mechanism. 

\subsection{Fermionic contributions} 

One can diagonalize $\Pi_0 (p^2)_{ij}$ and bring the fermion fields to have a canonical kinetic terms. For diagonal $\Pi_0 (p^2)_{ij}= \lambda_i^2 \delta_{ij}$, the scaling is simply $\psi_i \to \psi_i/\lambda_i$.  

We can separate the contributions to the matrix $\Pi^1_{ij}/\Pi^0_{ij} =( \Pi^1/\Pi^0)_{ij}$ for $i=j$ and $i\neq j$,  $$\Pi^1_{ij}/\Pi^0_{ij}= d_i \delta_{ij} + m_{ij} (1-\delta_{ij}) , \label{defd} $$ where no sum is implied and $\delta_{ij}$ is the Kronecker delta symbol.
This leads to a Coleman-Weinberg potential
\bea
V(\Phi) = - 2 \, \int \frac{d^4 p}{(2 \pi)^4} \left( \sum_i  \log \left( 1+ d_i \right) + \sum_{ij} \log \left( 1-\frac{1}{p^2} \, \frac{m_{ij}^2}{d_j d_i} \, \frac{\Pi^0_{i} \Pi^0_{j}}{\Pi^0_{ij}}\right) \right) \label{VphiPi}
\eea

Note that, in general $d_i$ and $m_{ij}$ will contain $s_\Phi$ and $c_{\Phi}$ terms. Also, the form factors $\Pi(Q^2)$ need to satisfy a set of Weinberg sum rules to render the integral finite, i.e. $\lim_{Q^2\to \infty} Q^{2 n}\Pi(Q^2)=0$. Whether these conditions are satisfied depends on the realization of the breaking. A common assumption is that the global symmetry $G$ is a chiral symmetry of some UV fermionic sector. A new strong  interaction, often called Technicolor (TC), is felt by the UV fermions, and causes fermion bilinears to condense. This condensation triggers the breaking of the global symmetry $G$ to $H$. If the new strong force is of the type $SU(N_{TC})$, for large values of $N_{TC}$, one can write the form factors as a sum over an infinite set of resonances,
\bea
\Pi(Q^2) = \sum_{n=1}^\infty \frac{f_n^2}{Q^2+m_n^2} \ , \label{largeN}
\eea 
with $m_n$ and $f_n$ their mass and decay constant, respectively. The Weinberg sum rules impose relations among these resonance parameters. For example, in extra-dimensional duals of this model, these resonances are identified with the Kaluza-Klein resonances, and one can then show that the form factors $\Pi(Q^2)$ do satisfy an infinite number of Weinberg sum rules due to non-locality in extra-dimensions~\cite{me-interp-others}. 

Provided the Weinberg sum rules are satisfied, the logarithm term can be expanded. At leading and next-to-leading order,
\bea
V(\Phi) = \alpha_c c_\Phi + \alpha_s s_\Phi + \beta_c c^2_\Phi + \beta_s s^2_\Phi + \ldots \label{VphiCWfinal}
\eea
where
\bea
\alpha_{c,s} &=& - 2  \sum_i  \, \int \frac{d^4 p}{(2 \pi)^4} d_i^{c,s} \nonumber \\
\beta &=&  2   \sum_{i,j}   \int \frac{d^4 p}{(2 \pi)^4}  \frac{1}{p^2} \, \left( \frac{m_{ij}^2}{d_j d_i} \, \frac{\Pi^0_{i} \Pi^0_{j}}{\Pi^0_{ij}} - d_i  d_j \right)
\eea
where $i$, $j$ run over the Dirac fermions.

\subsection{Inflationary models inspired on the Higgs as a pseudo-Goldstone boson} 

There is an extensive literature on the Higgs as a pseudo-Goldstone boson. Popular examples of 4D set-ups are Little Higgs~\cite{LH} and Composite Higgs models~\cite{Composite-Higgs-orig}. Among those, the minimal Composite Higgs model (MCHM) is relatively simple and is based on the breaking $\text{SO}(5) \rightarrow \text{SO}(4)$, but more elaborated models can be built with larger symmetry groups~\cite{CHM-reviews}. 

In this section we introduce the structure of the MCHM and use it as a template for inflation instead of a candidate for Higgs phenomenology. We will then use cosmological data to obtain information about the structure of the UV model, namely to reconstruct the shape of the form factors $\Pi(Q^2)$ required to generate inflation. This is meant to serve as an illustration of the mechanism and it is by no means the only way to write a successful inflationary model. Inspecting Eq.~\ref{VphiCWfinal}, one sees that a variety of potentials involving periodic functions arise from different breaking patterns.

In the MCHM, right-handed and left-handed third generation fermions (top and bottom quarks) contribute to the potential and gauge bosons interactions gauge a sub-group of  $\text{SO}(4)$.  The spinorial representation of  $\text{SO}(5)$ is as follows,
\bea \Gamma^{\hat{a}} = \left[ \begin{array}{cc} 0 & \sigma^{\hat{a}}  \\ \sigma^{\hat{a} \dagger} & 0 \end{array} \right], \,\,\,  \Gamma^{5} = \left[ \begin{array}{cc} \bf{1} & 0 \\ 0 &  -\bf{1}  \end{array} \right]   \textrm{ where } \sigma^{\hat{a}} = \{ \vec{\sigma}, -i \bf{1} \}   \eea

In this scenario, not all the possible form factors are generated. Indeed, if we inspect Eq.~\ref{VphiPi} and compare with the MCHM, the following relations are obtained, 
\bea \label{MCHM}
d^s_i =0 , \,  m_{ij}^c = 0 \ , 
\eea
and the resulting potential is of the form
\bea
V = \alpha c_\Phi - \beta s^2_\Phi \ .
\eea
The relation between the parameters $\alpha$, $\beta$ and the resonances has been fleshed out in the previous section. If we consider contributions to the inflaton coming from vector resonances and $N_F$ fermion flavors, one obtains~\cite{CHM-reviews}
\bea
\frac{\beta}{\alpha} = - \frac{9}{16 N_F} \frac{d_V}{d_F} \label{eqrat}
\eea
where $d$ is the ratio of form factors defined in Eq.~\ref{defd}, and the subscript $V$ ($F$) means vector (fermion) contributions and $N_F$ is the number of Dirac fermions.

In the next section we discuss the viability of this potential as a candidate for Natural Inflation and find that a specific relation between $\alpha$ and $\beta$ is required. The interpretation in terms of resonances of a strongly coupled sector responsible for the breaking $G/H$ can be done by noting that the ratio in  Eq.~\ref{eqrat} is related to a sum over bosonic and fermionic resonances.

\subsection{Inflating with a Higgslike PNGB}
We will consider inflation in the case (Eq.~\ref{MCHM}), that is, a potential of the form
\bea V ( \phi )  = \alpha \cos \frac{\phi}{f}  - \beta \sin^2 \frac{\phi}{f} \label{sincos} \eea

\begin{figure}[t] 
\centering
  \includegraphics[width= 200pt]{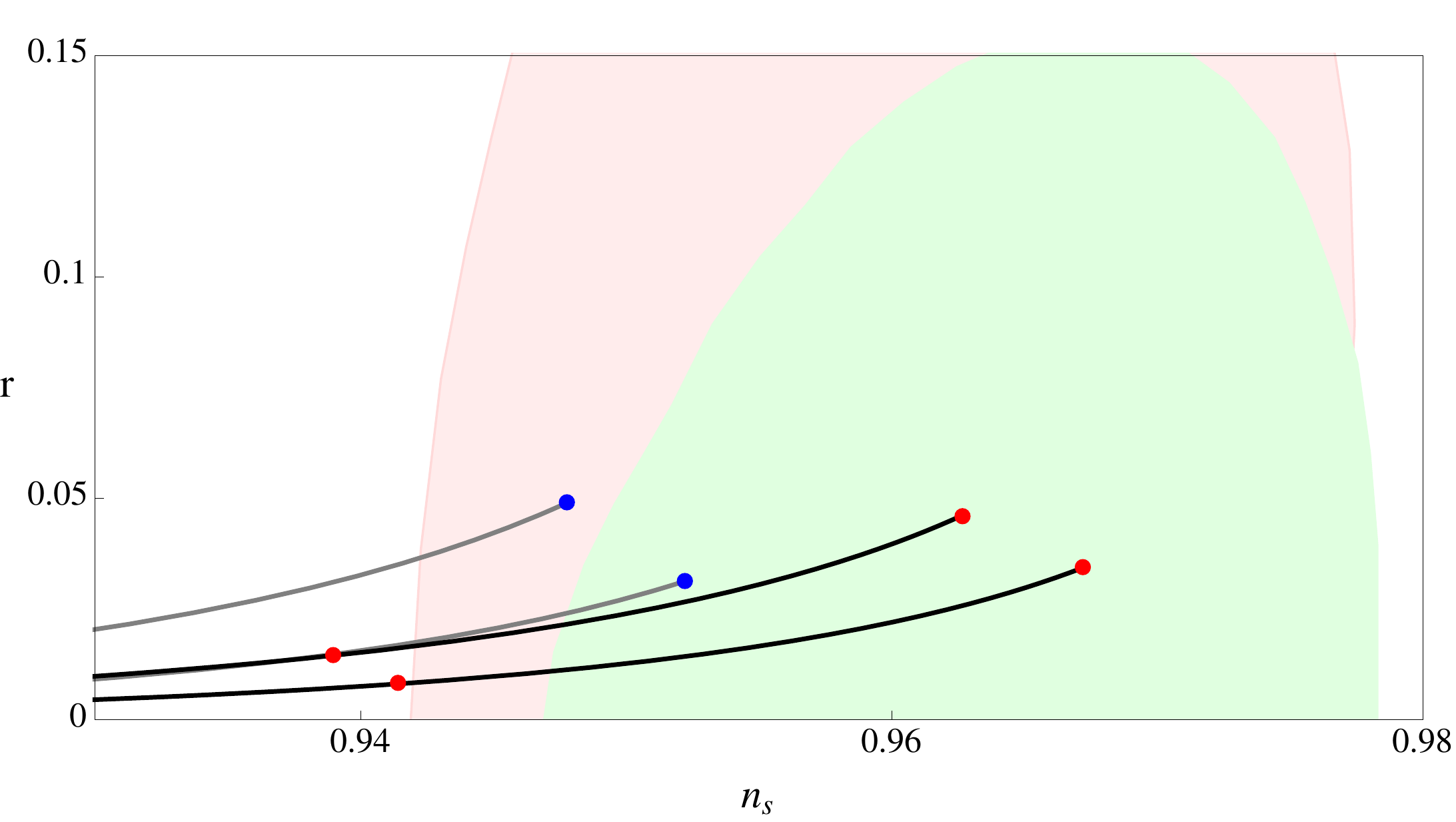}
  \caption{{\it Composite Higgs-like Inflation:} In grey $V(\phi) = \sin^2 \phi /f$, in black $V(\phi) = \cos \phi /f$ for $N=60 (50)$ for the lower (upper) line. The points $f = 10$ and $f = 7$ are marked. }
\end{figure} 
 
It is clear that in the case $\alpha = 0$ ($\beta = 0$) the potential reduces to a simple $\sin^2  \phi/f$ ($\cos \phi/f$) potential. As discussed in depth for $\cos \phi/f$ and shown in the figure below for $\sin^2  \phi/f$, this can only be compatible with slow-roll if $f > M_p$.

Therefore we will explore the range of potentials with both $\alpha$, $\beta \neq 0$ as it would be the case in models where the explicit breaking of the global symmetry is carried by both gauge and Yukawa interactions. We re-parametrize the potential as
\bea V ( \phi )  = \Lambda^4 \left(  \cos \frac{\phi}{f}  - \tilde\beta \sin^2 \frac{\phi}{f} \right) \ ,\eea
where $\tilde{\beta} = \beta/\alpha$ in Eq.~\ref{sincos}.

\begin{figure}[H] \label{wrongpot}
\begin{minipage}{.45\textwidth}
  \centering
  \includegraphics[width= 200pt]{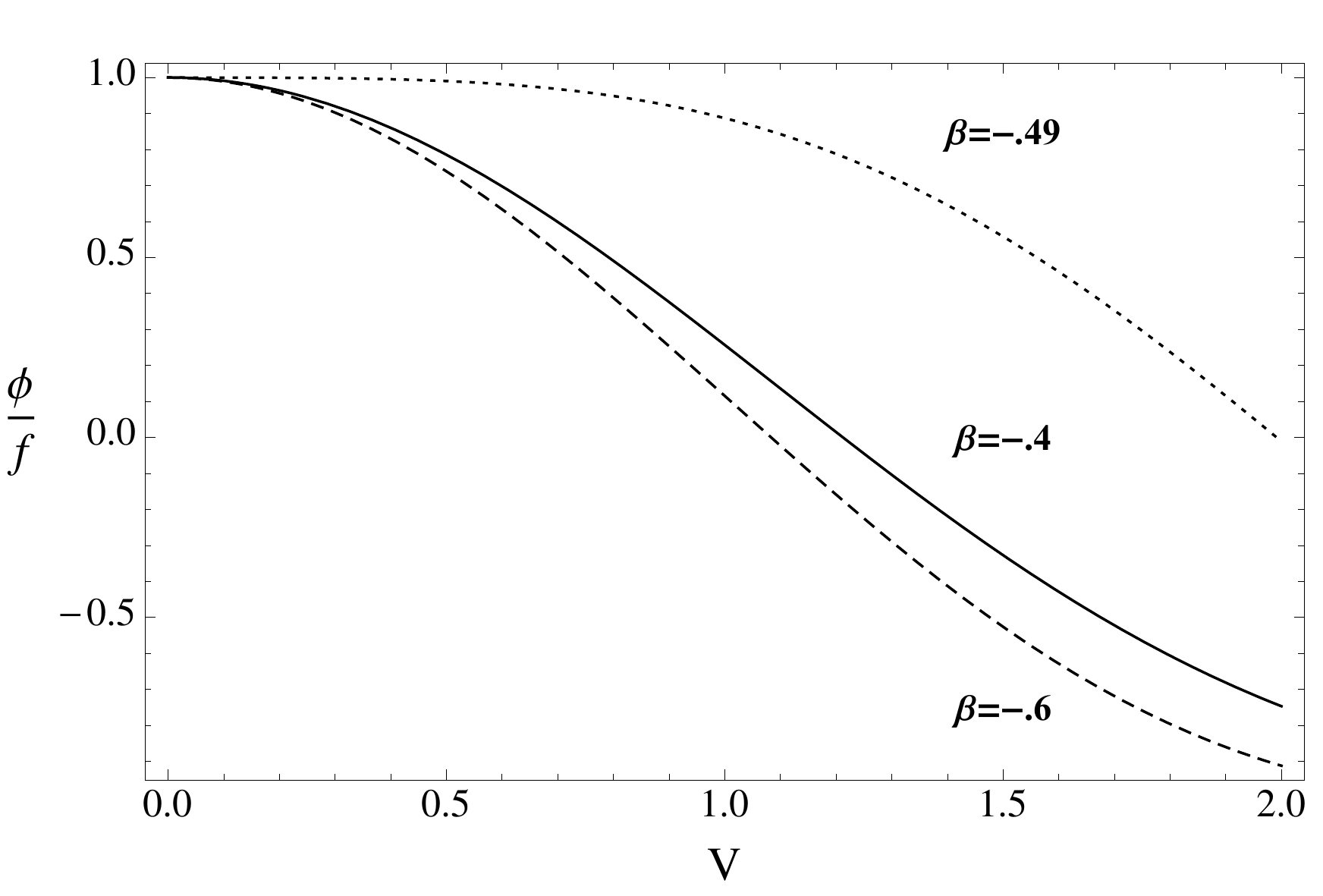}
  \caption{{\it Composite Higgs-like Inflation:} Shape of the potential for $f=1 \tilde{M}_p$ and $\tilde\beta = -0.49$, $\tilde\beta = -0.495$, $\tilde\beta = -0.497$} \label{potentialCW}
\end{minipage}
\hfill
\begin{minipage}{.45\textwidth}
  \hspace*{-1cm}\includegraphics[width=200pt]{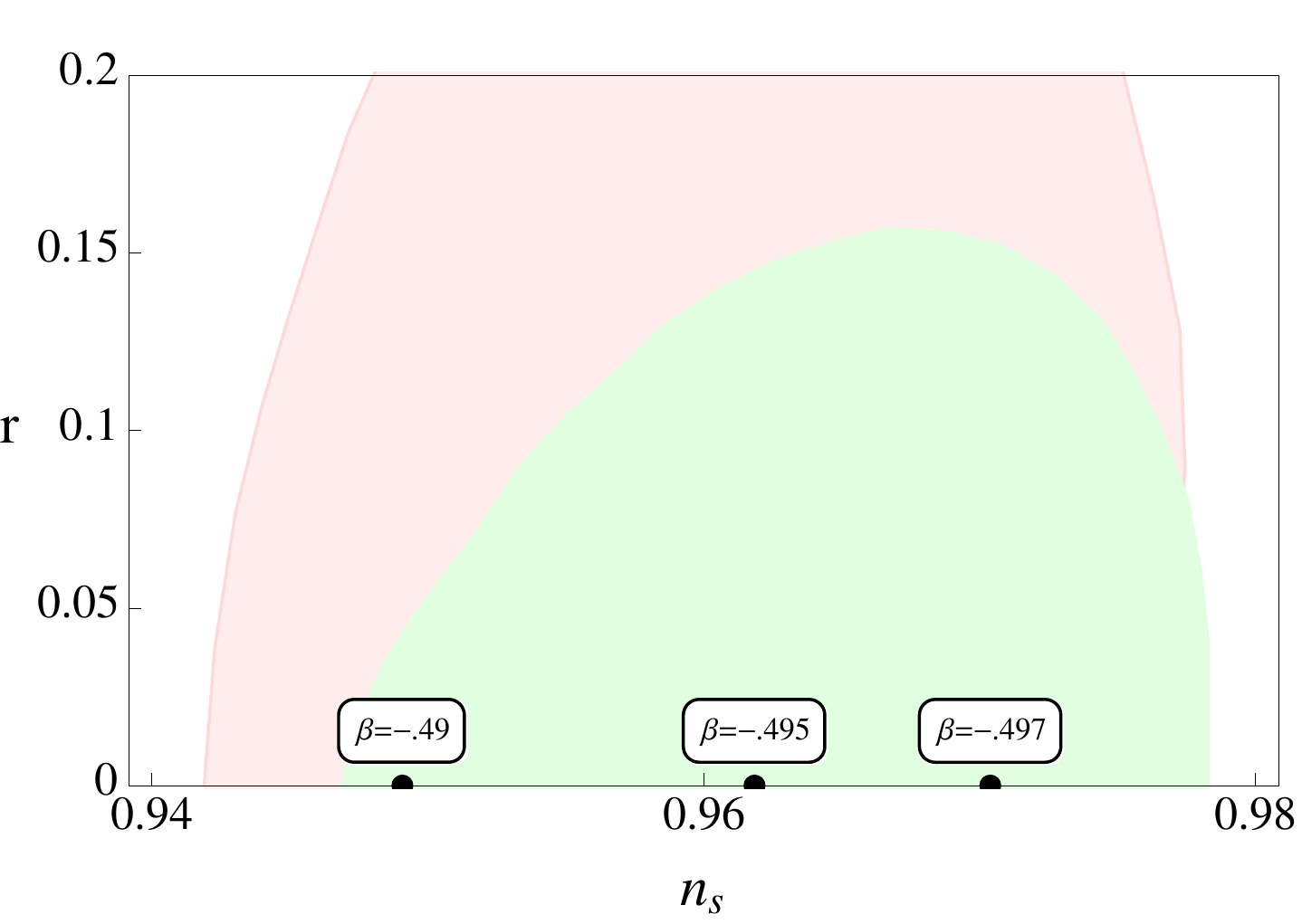}
  \caption{{\it Composite Higgs-like Inflation:} Predictions for $\tilde\beta = -0.49$, $\tilde\beta = -0.495$, $\tilde\beta = -0.497$ in the $n_s$-$r$ plane }\label{CWnsr}
\end{minipage}
\end{figure}

The potential has a very flat region around the origin for $- 1/2 \lesssim \tilde\beta < 0$ as can be seen from Fig.~\ref{potentialCW}. This range of $\tilde \beta$  can be translated into a region in the resonance parameter space using Eq.~\ref{eqrat}. In turn, it indicates a  specific relation between the vector and fermionic masses and decay constants, using Eq.~\ref{largeN}:
\bea
\tilde \beta = -1/2 \Rightarrow d_V = \frac{32}{9} N_F d_F \ .
\eea
At first sight, this relation may seem as unnatural,  a partial conspiracy among the form factors of spin-one and spin one-half particles in the resonance sector. Yet these kind of relations among form factors can be obtained in models with extra symmetries. For example, these ideas were used to build technicolor/higgsless models consistent with electroweak precision tests, where a cancellation between bosonic resonances~\cite{HTC} or fermionic and bosonic resonances~\cite{CuredHless} was achieved in a walking theory.

\begin{figure}[t] 
\centering
  \includegraphics[width= 200pt]{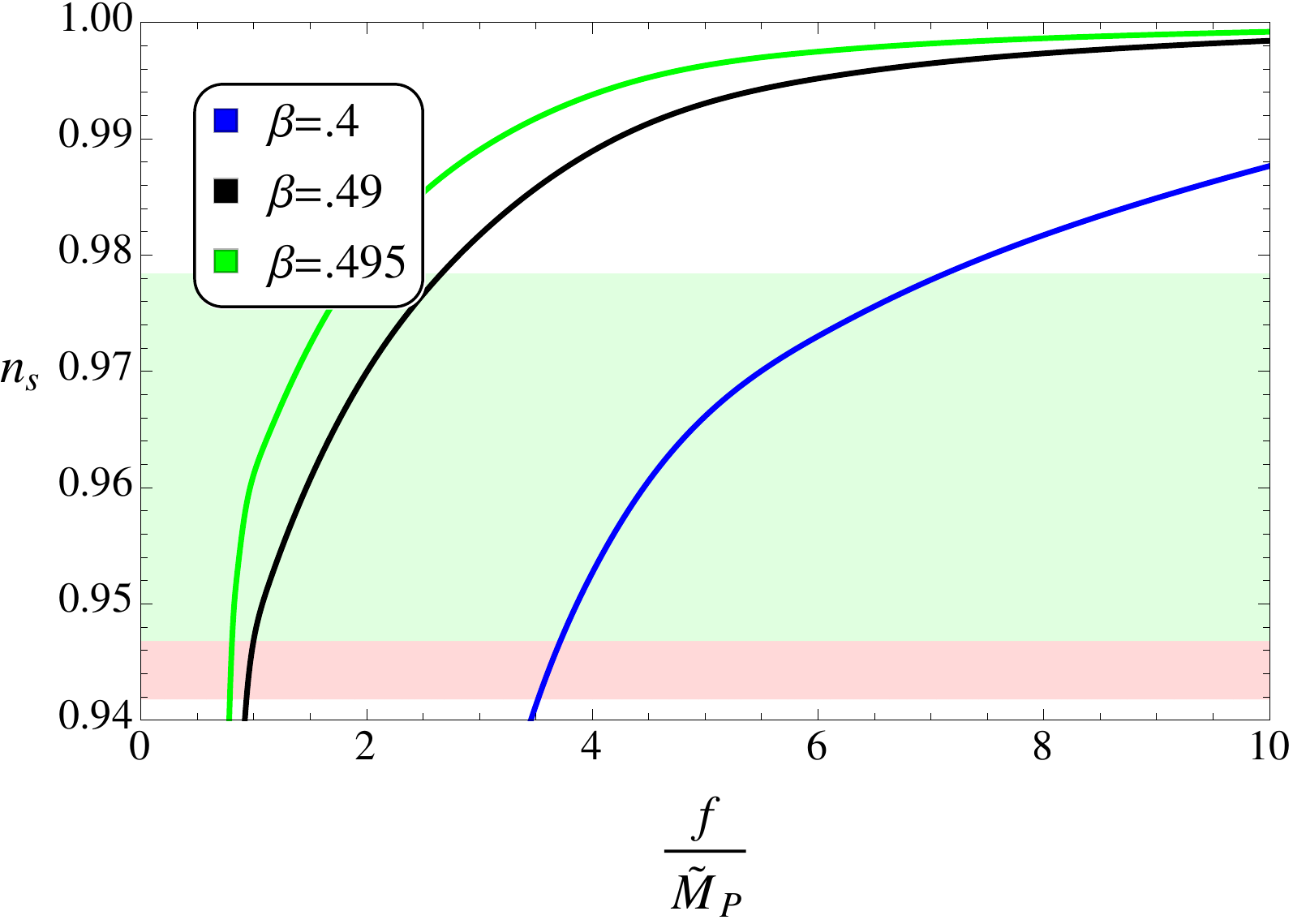}
  \caption{{\it Composite Higgs-like Inflation:} Predictions for $\tilde\beta = -0.4$, $\tilde\beta = -0.49$, $\tilde\beta = -0.495$ in the $f$-$n_s$ plane.  }\label{CWnsf}
\end{figure}

This potential can indeed lead to sub-Planckian values for $f$ and agree with Planck data. For example, we show the region for $f=\tilde{M}_P$ in figure \ref{CWnsr} the sensitivity to the value of $\tilde \beta$. In Fig.~\ref{CWnsf} we show that for $\tilde \beta $ close to -1/2, $f <  \tilde{M}_p$ and satisfy Planck's constraints on $n_s$.

In the cases we consider, $r \ll \mathcal{O}(1)$ (as sees clearly figure \ref{CWnsr}); this remains the case for different values of $f$. In figure \ref{CWnsf} we compare predictions for $n_s$ with different $f$. 

To second order, the potential is
$$ V (\phi) = \Lambda^4 \left( \cos \frac{\phi }{f}- \tilde\alpha_2 \cos ^2\frac{\phi }{f}+ \tilde\beta \sin ^2\frac{\phi }{f} +  \tilde\beta_2 \sin ^2\frac{\phi }{f} \left( 1 + \cos \frac{2 \phi }{f} \right)\right)$$

where $$\tilde\alpha_2 = \frac{\alpha_q^2+\alpha_u^2 }{2 (\alpha_q-\alpha_u)} =\frac{\alpha_q^2+\alpha_u^2}{2 \alpha}   \,\, , \,\,\,\,\,\,\,\,\,\,\,\,\,\,\,\,\,\,\,\,\,\,\tilde\beta_2 = \frac{\beta_f^2 +\beta_g^2}{4 \alpha} $$

One then expects $\tilde\alpha_2$ and $\tilde\beta_2$ to be small, although larger if  $\alpha_q$ and $\alpha_u$ are very close to each other.

\section{The effect of UV completions: Higher-Dimensional Terms}

Up to now we have discussed generation of the inflationary potential through infra-red (IR) effects, such as non-perturbative instanton, a Wilson line or a Coleman-Weinberg potential. As we discussed in the introduction, when $f>\tilde{M}_P$, the effect of UV corrections of the type in Eq.~\ref{NRpot} is potentially dangerous. These would generate higher-dimensional operators (HDOs) which, for large values of the field, could dominate over the renormalizable terms. In this section we discuss the effects of non-renormalizable terms in the case of the Vanilla Natural Inflation discussed in Sec.~\ref{vanilla} to show how sensitive these models are to HDOs in the regime of $f> \tilde{M}_p$. 

We already mentioned wormhole terms in the Introduction, but there are other possible origins for HDOs. For example, graviton loop corrections~\cite{ignoble, smolin} would preserve the cosine form of the potential
$$ \delta V = M^4 c_n \left[ \cos \frac{n \phi}{f} \right]^m . $$
By choosing $m$ and $n$ appropriately, one can account for the terms we invoked  in the Introduction, Eq.~\ref{HDOs}. 

Also higher-order instanton effects could influence the HDOs. As explained in Sec.~\ref{vanilla}, instanton effects would maintain the periodicity of the potential.
They thus will be of the form ($n > 1$)\footnote{Note that this is the potential of Extra-natural Inflation, in which there is a sum over $n$. The term $n=1$ gives the natural inflation model.}
\bea \delta V = \Lambda^4 \tilde{c}_n \cos \frac{n \phi}{f} = \Lambda^4 \tilde{c}_n \sum_{k=0}^{\infty} \frac{(-1)^k \left(\frac{n \phi }{f} \right)^{2 k}}{(2 k)!} = \Lambda^4 c_n \sum_{k=0}^{\infty} (-1)^k  \phi ^{2 k} \eea
Of course, the expanded form masks the underlying shift symmetry, but it allows us to connect with the HDO language. 
It is clear that the coefficients $c_n$ are growing with $n$, and will thus ruin slow-roll inflation unless the coefficients $\tilde{c}_n$ decrease with $n$. The extra terms will not influence the slow-roll predictions from Natural Inflation much in the limits
\bea \tilde{c}_n  n^2 \ll 1  \textrm{ and } \frac{\phi}{f} \lesssim 2 \eea
This is the case in Extra-natural inflation, in which $\tilde{c}_n \propto 1/n^5$. For non gravitational effects, $\tilde{c}_n$ decreases with $n$ as $e^{-n S}$ where $n$ is the number of instantons ~\cite{ignoble}. 

Finally, UV breaking of the inflaton's shift symmetry such as new fermion masses could generate new HDOs.

\subsection{Adding Higher-Dimensional Operators to Vanilla Natural Inflation}

Here we consider the effects of higher order operators added to the vanilla Natural Inflation model. We parametrise the effect as
\bea V( \phi ) =  \Lambda ^4 \left[1 + \cos \left(\frac{\phi }{f}\right) +  \sum_{n=5}^{n=8} c_n \, \phi^n  \right] \eea
where $c_n$ are dimensionless, as we have rescaled all dimensionful parameters by $\tilde{M}_p$. These non-renormalizable terms are higher-dimensional operators (HDOs). Note that the sensitivity of models with values of the field larger or close to the UV cutoff  has  been recently studied in, e.g. Refs.~\cite{xavier,valerie} for quantum gravity and SUSY unification HDOs. 

Since we assume that the HDO will give corrections to the leading order model, we solve for $\phi_N$ in this model perturbatively,
\bea \phi_N = \phi_{NI} + \epsilon \, \phi_1 \eea
using $ N (  \phi_{NI} + \epsilon \phi_1 ) =60 $. The zeroth order equation will be solved by (Eq.~\ref{phiNI}).

\begin{figure}[H] 
\centering
  \includegraphics[width= 300pt]{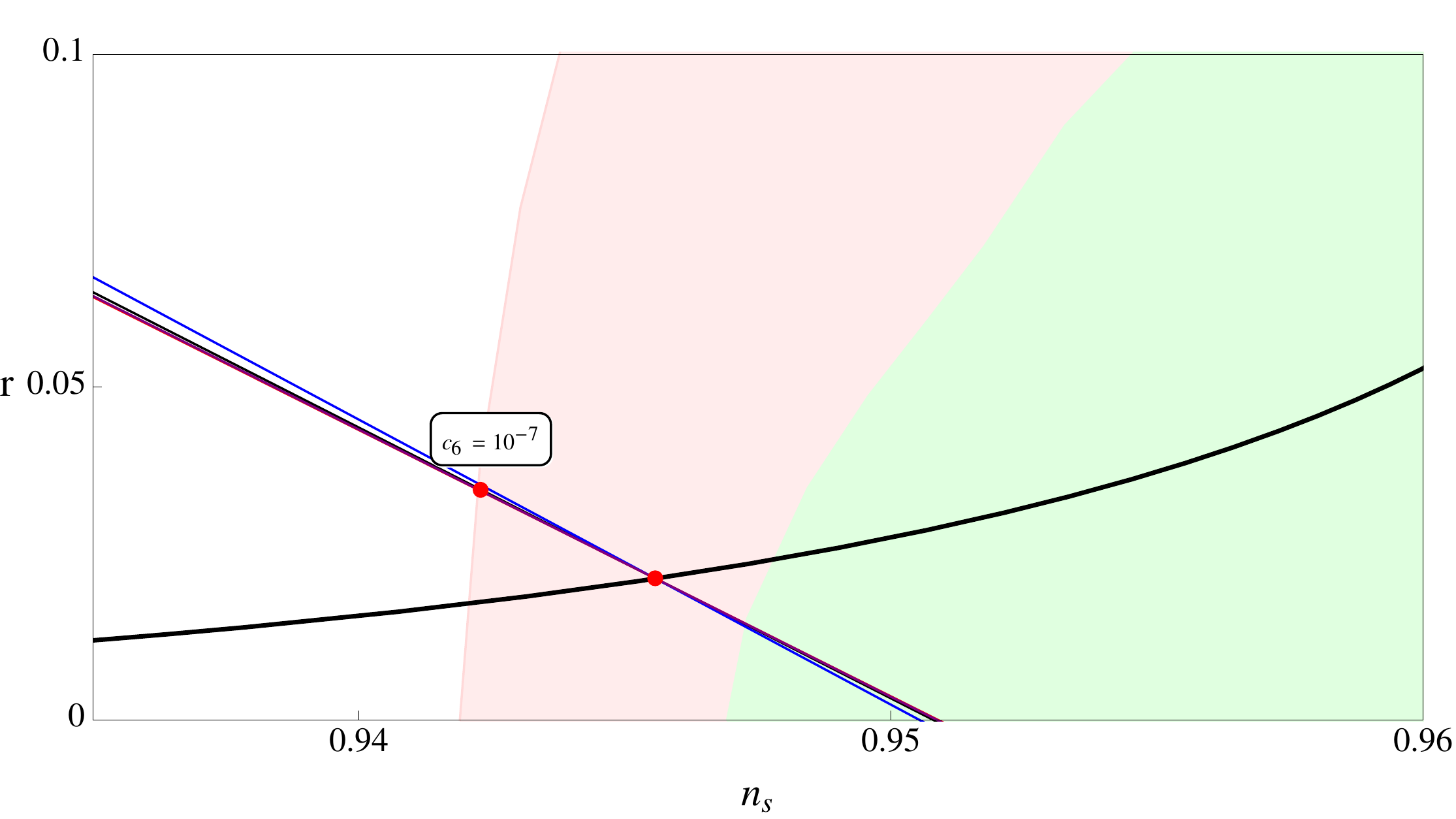}
   \caption{{\it Vanilla Natural Inflation with non-renormalizable corrections: }Varying the coefficients of HDOs in the $n_s-r$ plane for $f \simeq 4.5 \tilde{M}_p$. With $c_5$, $c_6$, $c_7$, $c_8$ in blue, black, purple, and black respectively. At $r = 0$, $ \Delta \phi \sim \phi_I = 0.21 M_p$, $|\xi_V | < 4 \times 10^{-5}$ and $ \varpi_V < 8 \times 10^{-7}$. At the Natural inflation reference point ($c_n= 0$, $\forall n$),  $ \Delta \phi \sim \phi_I = 0.24 M_p$. The point at the edge of the pink region corresponds to $c_6$= $10^{-7}$. }\label{vanillaHDO}
\end{figure}

In figure \ref{vanillaHDO} we show the variation of the coefficients $c_n$ (where $c_m =0$ for $ m \neq n$) in the $n_s-r$ plane for $f=0.9 M_p \simeq 4.5 \tilde{M}_p$.   All HDOs give contributions in a similar direction in the $n_s$, $r$ plane. 

The effect of tiny values of $c_n$ could bring the model outside the Planck region. For example, the introduction of a small $\phi^6$ term with a $c_6\sim 10^{-7}$ does bring the model to unacceptable values of $n_s$ and $r$. This is an illustration of how the predictions of inflationary models in the trans-Planckian region are questionable.

Let us finish with some comments on the effect of HDOs in this model. Solving for the values of the dimensionless constants $c_n$ at $r=0$ ($c_m = 0$, $m\neq n$), we have
\begin{center}
  \begin{tabular}{| l | r |}
    \hline
     & r = 0  \\ \hline
    $ c_5 $ & $- 1.85 \times 10^{-6}$ \\ \hline
    $ c_6 $ & $- 1.58 \times 10^{-7}$ \\ \hline
    $ c_7 $ & $- 1.32 \times 10^{-8}$  \\ \hline
    $ c_8 $ & $- 1.10 \times 10^{-9}$  \\ \hline
  \end{tabular}
\end{center}
To make the perturbed natural inflation more compatible with the CMB data, the coefficients need to be small and negative, although the statement can be made less strong if a cancelation between them occurs. 

For different values of $f$ the contributions by the HDOs in the $n_s$, $r$ plane will be similar. This allows us to find a lower bound on $f$ for which Natural Inflation can be made compatible with the Planck data (the blue region in plot (Eq.~\ref{NIplot}), as Natural Inflation predicts no running) with the aid of HDOs: 
$$ f \gtrsim 0.868 M_p = 4.35 \tilde{M}_p .$$

\begin{figure}[H] 
\centering
  \includegraphics[width= 200pt]{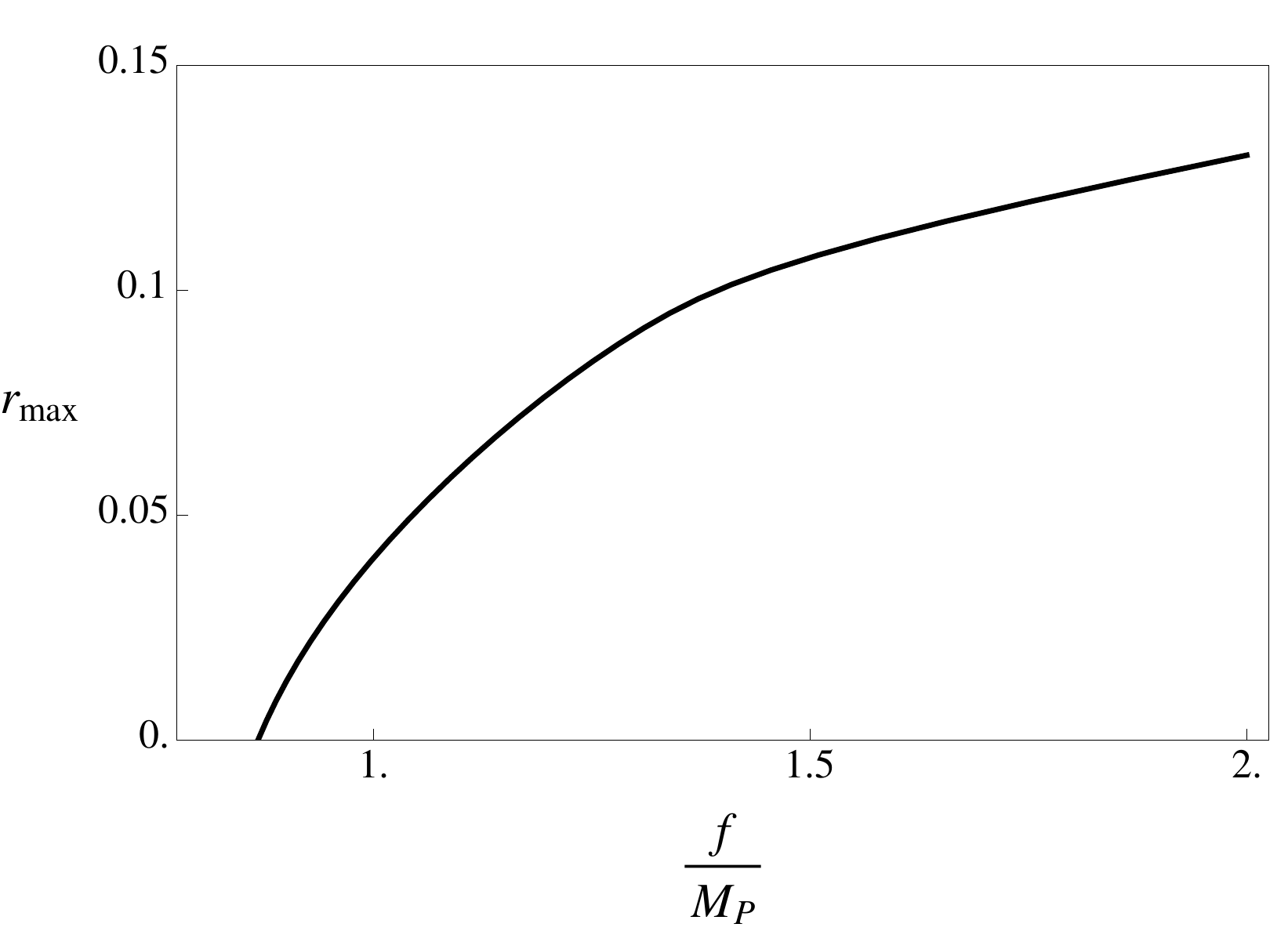}
  \caption{Maximal tensor to scalar ratio $r$ as a function of $f$ in vanilla Natural Inflation models and the operator $c_6 \phi^6$.}
\end{figure}

\section{Conclusions}

The paradigm of slow-roll inflation after the Planck data, albeit very successful parametrization, requires a questionable field theory approach. The inflationary hierarchy problem, namely the  tension between providing sufficient inflation yet satisfy the amplitude of the CMB anisotropy measurements, implies that the width of the potential must be much larger than its height. This tuning is generically unstable unless some symmetry protects the form of the potential. This is the idea of Natural Inflation~\cite{here}:  the inflationary potential is protected by a shift symmetry provided the inflaton is a pseudo-Goldstone boson.  

The archetypical example of Natural Inflation is an axion-like inflaton with a potential generated by instantons. This specific form of the potential requires large values of the inflaton field to sustain sufficient inflation. Extra-natural inflation, based on an extra-dimensional realization of the same idea, suffers from  the same problem. The unknown higher order corrections could easily upset the inflationary model predictions, implying that these models are not a good effective theory.

In this paper we aim at preserving the idea of Natural Inflation as a solution to the cosmological hierarchy problem in models with an adequate effective description, where the field does not undertakes any trans-Planckian excursion. We find that there are many of such models once we go beyond the original Natural Inflation model. 

Indeed, in this paper we have explained how Extra-natural inflation can be generalized to encompass (more realistic) extra-dimensional models with bulk fermions. A large number of such bulk configurations do lead to a successful inflationary model with $f > \tilde M_{P}$ and a moderate amount of bulk matter content.

We have shown how four-dimensional models with Higgs-like potentials can also lead to successful inflation. We have presented the study of a specific such model based on the Minimal Composite Higgs model, but also developed a formalism to explore different breaking scenarios. We also showed how, once the inflationary conditions are set, one can obtain information about the sector of heavy resonances which accompanies the pseudo-Goldstone boson in these scenarios.
    
In the case of Vanilla Natural Inflation, we have studied the effect of higher-dimensional operators. In this case, as $f > \tilde{M}_p$, these higher-dimensional effects gives us a sense of how sensitive is the theory to the UV completion. We have found that moderate values of the operators do change drastically the cosmological predictions~\cite{xavier,valerie}, emphasizing once more the necessity of sub-Planckian field values to retain the predicitivity of the model.

\section*{Acknowledgements}
We would like to thank Xavier Calmet and Daniel Litim for discussions about quantum gravity corrections. The authors would like to acknowledge funding from grant STFC ST/L000504/1.

 \providecommand{\href}[2]{#2}\begingroup\raggedright\endgroup

\end{document}